\documentclass[amsmath,superscriptaddress,prd]{revtex4}

\usepackage{graphicx}

\newcommand{\lpa}{\hbox to0pt{\raise7pt\hbox{$\leftarrow$}\hss}\partial}
\newcommand{\slD}{\hbox{$D\kern-7pt/$}}
\newcommand{\slx}{\hbox{$x\kern-6pt/$}}

\begin{document}
\title{ Improved Determination of the  Mass of the $1^{-+}$ Light Hybrid Meson
From QCD Sum Rules}
\author {H.Y.  Jin}
\affiliation{Institute of Modern Physics, Zhejiang University, P. R. China
 }
\affiliation{Department of Physics \& Engineering Physics, University of 
Saskatchewan, Canada}
\author{J.G. K\"orner} 
\affiliation{Institut f\"ur Physik, Johannes Gutenberg-Universit\"at,
 Germany }
\author{T.G. Steele}
\affiliation{Department of Physics \& Engineering Physics, University of 
Saskatchewan, Canada}

\begin{abstract}
We calculate the next-to-leading order (NLO) $\alpha_s$-corrections to the contributions of  the 
condensates 
$\langle \alpha G^2\rangle$ and $\langle \bar qq\rangle^2$ in 
the current-current 
correlator of the hybrid current  $g\bar
q(x)\gamma_{\nu}iF_{\mu\nu}^aT^aq(x)$ using the external field method
in Feynman gauge.  After incorporating these NLO contributions into the Laplace sum-rules, 
 the mass of 
the $J^{PC}$=$1^{-+}$ light hybrid meson is recalculated using the QCD sum rule approach.
We find that the sum rules  exhibit enhanced stability when the NLO $\alpha_s$-corrections
are included in the sum rule analysis, resulting 
in a $1^{-+}$ light 
hybrid meson mass of approximately  1.6 GeV. 
\end{abstract}

\maketitle

\section{Introduction}
 Soon after Quantum Chromodynamics (QCD) became established as the theory
of strong interactions, the search for mesons with exotic $J^{PC}$
quantum numbers was initiated. 
Important experimental progress has occurred in identifying
potential candidates for exotic mesons. Among these are the 
two  $J^{PC}=1^{-+}$ isovector states $\hat\rho(1405)$ and  $\hat\rho(1600)$  
which have been identified  by  two collaborations
 in the channels $\rho\pi$, $\eta\pi$ and 
$\eta^\prime\pi$ \cite{bnl1,bnl2,bnl3}. This gives the theorists an opportunity to compare their 
results for the exotic light hybrid mesons with the experimental observations.  For instance,
the widely cited mass prediction for the  $J^{PC}=1^{-+}$ state from lattice  
QCD lies around 1.9 GeV \cite{lattice} 
which disagrees with the  
experimental data. The decays of the $1^{-+}$ hybrid meson have also been  
studied in the context of various models \cite{page,ailin}, and also 
appear to be in
disagreement with the experimental data. A
possible reason for these inconsistencies may derive from the fact that 
non-perturbative effects cannot be easily controlled at this low energy 
scale. A further possibility is, of course, that the two new states may have been
misidentified and are not hybrid mesons after all. 
It is clear that further theoretical studies of the properties of the hybrid
mesons are necessary before one can be confident of their discovery. 

In this paper, we concentrate on the mass prediction for the $1^{-+}$  
hybrid meson using the QCD Laplace sum rule approach  \cite{shif}. 
Previous theoretical studies found masses in the range  1.4--2.1~GeV \cite{lead}, with more recent 
estimates around 1.6 GeV   
\cite{narison,jin}. These latter estimates are close to the  $\hat\rho(1600)$  
 but do not accommodate the state $\hat\rho(1405)$ 
as a hybrid candidate. However, the QCD sum rule analysis is not very stable,
which from sum-rule studies of scalar gluonium and the $\rho$ meson, could be
attributed to the importance of NLO corrections associated with the dimension-four  
 $\langle\alpha_sG^2\rangle$ and dimension-six $\langle\bar qq\rangle^2$ operators
\cite{chen,glue}.
In particular,
the four-quark operator has the same dimension as the two point correlator 
of the $1^{-+}$ current and its coefficient function 
is non-logarithmic in the leading order. Thus the dimension-six contributions are absent 
in the next-to-lowest moment sum-rule, resulting in discrepancies 
with the mass prediction from  
the zeroth moment sum rule  \cite{lead}.  
Thus it  is
 interesting to determine  whether the $\alpha_s$-corrections can reduce this discrepancy.   
For consistency, the $\alpha_s$-corrections to both the dimension-four  
$\langle\alpha_sG^2\rangle$ and the dimension-six $\langle\bar q q\rangle^2$ operators need to be included. 

In this paper, 
we first give a brief introduction to the external field method  
used for our calculations. We then present our calculations for 
the $\alpha_s$ corrections to $\langle\bar qq\rangle^2$ and
$\langle\alpha_sG^2\rangle$. Finally, we discuss the effect of
the $\alpha_s$-corrections on the mass prediction for the $1^{-+}$ hybrid meson.

\section{External field method}\label{ext_field_section}
In order to obtain the contribution of the operator 
$\alpha_sG^{a\mu\nu}G^a_{\mu\nu}$
in the Operator Product Expansion (OPE) expansion of the relevant current-current
correlator, it is convenient to use
the so-called external field method introduced in 
Ref. \cite{shif1}. Because we will use a slightly modified version of the external
field method, we give a
brief introduction to this method  before carrying out our 
calculations. We first split the gauge field
$A_\mu^a$ into two parts 
\begin{equation}\label{field}
A_\mu^a = (A_\mu^a)_{ext} +a_\mu^a,
\end{equation}
where $(A_\mu^a)_{ext}$ is an external (classical) field and $a_\mu^a$ is a 
quantum field. In the following we will suppress the label ``ext" in $(A_\mu^a)_{ext}$ 
for convenience. Substituting (\ref{field}) in the QCD Lagrangian one obtains 
\begin{equation}\label{lang}
L=-\frac{1}{4}F^{a\mu\nu}F^a_{\mu\nu}
+\bar\psi i\slD\psi
\end{equation}
where 
\begin{equation}
F_{\mu\nu}={\frac{1}{ig}[D_\mu,D_\nu]}
\quad ,\quad D_\mu=\partial_\mu + iga_\mu+ igA_\mu
\end{equation} 
The Lagrangian (\ref{lang}) is invariant under the transformation
\begin{gather}\label{trans}
a_\mu(x) \rightarrow a'_\mu = U^{-1}(x)(a_\mu(x)+A_\mu(x))U(x)-A_\mu(x)
+\frac{i}{g}U^{-1}(x)U(x),\\
\psi(x) \rightarrow \psi' = U(x)\psi(x),\\
A_\mu \rightarrow A^\prime_\mu(x)=A_\mu(x)\quad .
\end{gather}
Unlike in \cite{shif1}, we demand that the external field is invariant 
under the gauge transformation. Therefore, any gauge condition
imposed on the external field does not break the invariance under
the gauge transformation (\ref{trans}). This is consistent with 
the result given in \cite{ste2}, which states that the series of the 
OPE for a correlator of gauge invariant 
currents does not depend on the choice the gauge of the external
field. Furthermore, in order to fix the gauge of the quantum
field, we use the Feynman gauge 
\begin{equation}\label{gauge}
{-\frac{1}{2}(\partial_\mu a^{\mu})^2}
\end{equation}
instead of the background gauge $-\frac{1}{2}(D^{ext}_\mu a^{\mu})^2$ used
in \cite{shif1},
 where $D^{ext}_\mu=\partial_\mu+igA_\mu$. This has the advantage that it can be
shown that the external method in the covariant gauge is 
equivalent to the plane-wave method \cite{ste2}. 
Note that, when the external field method is used, the
background gauge very likely differs from the covariant gauge in 
processes where both the radiation field
 and the  external
field are present  in the initial states and (or) the  final states. 
In the processes where
only  the radiation field or the external field  is present in the initial 
states and (or) the  final states, there is no difference between  the 
background gauge and the covariant gauge. For instance, if there is 
only the external 
field present in the initial and (or) final states, the radiation field 
is integrated out. Then the  external method in the background gauge is
exactly the same as the original background gauge method \cite{back}. 
The later is consistent with the plane-wave method. If there is only the
radiation field present in the initial state and (or) final state, we can 
switch off the external field. Then there is no difference between the 
background gauge and the covariant gauge.    

The Feynman rules in the Feynman gauge can be obtained  quite straightforwardly.  
Under the infinitesimal gauge transformation 
\begin{equation}
\delta
a^a_\mu=-f^{abc}\omega^b(a^c_\mu+A^c_\mu)+\frac{1}{g}\partial_\mu\omega^a,
\end{equation}
one can easily derive the  Lagrangian for the ghost field 
\begin{equation}
L_{ghost}=-\theta^\dag_a[\partial^2\delta^{ab}-\lpa
f^{acb}(a^c_\mu+A^c_\mu)]\theta_b.
\end{equation}
Then,  the external field $A_\mu$ obeys the same Feynman rules as  
the  radiation field $a_\mu$. This is also consistent with  the 
plane-wave method \cite{ste2}.  
The calculational techniques using the Lagrangian (\ref{lang}) are 
similar to those in the background gauge \cite{back}.

In order to extract the operator $\alpha_sG^{a\mu\nu}G^a_{\mu\nu}$ from 
the OPE,
the Fock-Schwinger condition is imposed on the
external field 
\begin{equation}\label{schw}
x^\mu A^a_{\mu}(x)=0.
\end{equation}
Then, with the aid of the technique proposed in \cite{shif1},
the gluon propagator in the presence of the external field can be obtained
straightforwardly. For instance, up to the term ${\cal O}(q^{-4})$ the gluon
propagator is given by 
\begin{equation}\label{gp}
\begin{split}
D_{\mu\nu}(q,y)&=\int {\mathrm e}^{iq\cdot x}\mathrm{d}^Dx\,
\langle 0|Ta^a_\mu(x)a^b_\nu(y)|0\rangle\\
&=\frac{-i}{q^2}\left[\delta^{ab}g_{\mu\nu}-
\frac{3}{2}\frac{G^{ab}_{\mu\nu}}{q^2}+\frac{q^{\rho}(G^{ab}_{\rho\mu}
q_\nu+G^{ab}_{\rho\nu}q_\mu)}{q^4}-\frac{iy^\rho}{2}\frac{(G^{ab}_{\rho\mu}
q_\nu+G^{ab}_{\rho\nu}q_\mu)}{q^2}+
g_{\mu\nu}iy^\rho G^{ab}_{\rho\sigma}
\frac{q^\sigma}{q^2}\right],
\end{split}
\end{equation}
where $G^{ab}_{\mu\nu}=gf^{acb}G^c_{\mu\nu}$ and $D=4-2\epsilon$ is the dimension of
space-time. As expected, the gluon propagator (\ref{gp})  differs 
from the corresponding one given in \cite{shif1}. 
Sometimes it 
is more convenient to do the calculation in coordinate space. The gluon 
propagator in coordinate space can be obtained by using 
the $D$-dimensional  Fourier transformation 
\begin{equation}
D_{\mu\nu}(x,y)=\int \frac{\mathrm{d}^Dq}{(2\pi)^D} {\mathrm{e}}^{-iq\cdot x}D_{\mu\nu}(q,y)
\end{equation}
The necessary integration techniques in  $D$-dimension were given in 
\cite{shif1}. One can convert
 the 4-dimensional quark propagator given in
\cite{shif1} to the $D$-dimensional quark propagator 
\begin{equation}
S(x,0)={\frac{\Gamma(\frac{D}{2})}{2\pi^{\frac{D}{2}}}
\frac{\slx}{(x^2)^{\frac{D}{2}}}-\frac{\Gamma(\frac{D}{2}-1)}{8\pi^{\frac{D}{2}}}
\frac{x_\alpha{\tilde
G}^{\alpha\beta}\gamma^\beta\gamma_5}{(x^2)^{\frac{D}{2}-1}}}.
\end{equation}

\section{NLO $O(\alpha_s)$ corrections to $\langle G^2\rangle$ and  
$\langle\bar qq\rangle^2$}
The two point correlator of the $1^{-+}$ hybrid current is defined as 
\begin{equation}\label{cor}
\begin{split}
\Pi_{\mu\nu}(q^2)&=i\int \mathrm{d}^4x\, {\mathrm{e}}^{iqx} \langle 0|
T\{j^{ren}_{\mu}(x),j^{ren +}_{\nu}(0)\}|0 \rangle \\
&={(q_\mu q_\nu-g_{\mu\nu}q^2)\Pi_v(Q^2)+q_\mu
q_\nu\Pi_s(q^2)},
\end{split}
\end{equation}
where the current $j_{\mu}(x)=\bar q(x)T^a\gamma_{\nu}igF^a_{\mu\nu}q(x)$ carries 
isospin $I=1$ 
and the invariants
$\Pi_v(Q^2)$ and $\Pi_s(Q^2)$ correspond to the contributions from
$1^{-+}$ and $0^{++}$ states respectively.
The renormalized current is denoted by $j^{ren}_{\mu}(x)$, which 
in the massless 
quark limit, has the form 
\begin{equation}\label{z}
j^{ren}_\mu=Zj_\mu,
\end{equation}
where up to the order ${\cal O}(\alpha_s)$ and in the Feynman 
gauge the renormalization constant $Z$ is given by \cite{jin} 
\begin{equation}
Z={1+\frac{2}{9}\frac{g^2}{\pi^2}\frac{1}{\epsilon}}
\end{equation}

\begin{figure}[hbt]
\centering
\includegraphics[scale=0.12]{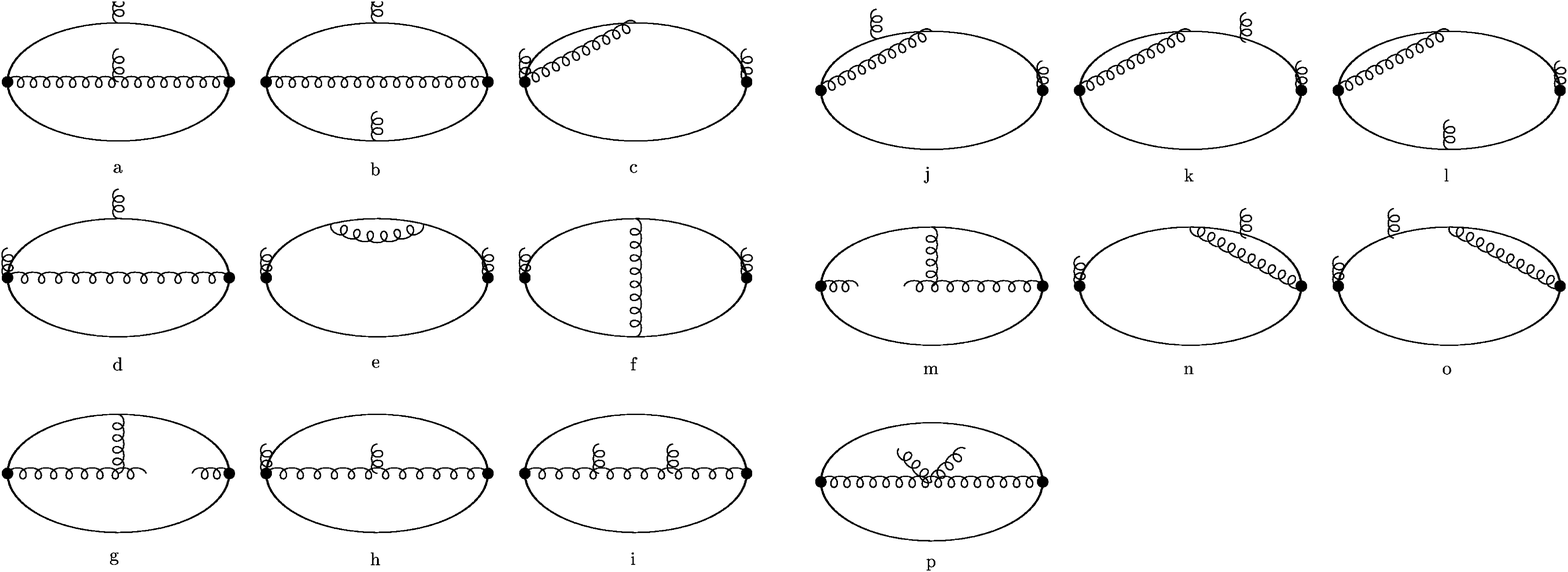}
\caption{
Feynman diagrams for the $\alpha_s$-corrections to the condensate  $\langle
\alpha_s G^2\rangle$. Dots 
stand for the current vertices.
}
\label{glue_fig}
\end{figure}

The leading order calculation of (\ref{cor})
including the quark and gluon condensate contributions are contained in 
\cite{lead}, and the NLO corrections to the
perturbative part of (\ref{cor}) were calculated in
\cite{narison,jin}. 
Next we  consider the  NLO corrections to  the gluonic
condensate $\langle\alpha_sG_{\mu\nu}G^{\mu\nu}\rangle$ contributions. 
We divide the calculations into two parts. 
One part can be obtained via the calculations of Feynman diagrams shown in
Figure \ref{glue_fig}, where we do not display diagrams which vanish in
dimensional regularization or which can be obtained from diagrams a--o
in Fig.\ \ref{glue_fig} by symmetry arguments. Because the expansion in term of 
$x_\mu$ violates translation invariance, the Fig.\ \ref{glue_fig} diagrams g, j, and k 
respectively differ from diagrams m, n, and o.  
 A straightforward calculation results in the structure
\begin{equation}
\begin{split}
\Pi^{G^2}_{\mu\nu}(q^2)= & 
\left(\frac{q_\mu q_\nu}{q^2}-g_{\mu\nu}\right)
\alpha_sG_{\alpha\beta}G^{\alpha\beta}\Pi^{G^2}_1(q^2)+
\left(\frac{q_\mu q_\nu}{q^2}-g_{\mu\nu}\right)\alpha_sG_{\rho\beta}G_\sigma^\beta
q^\rho q^\sigma \Pi^{G^2}_2(q^2)\\
&+
\left(\alpha_sG_{\mu\beta}G_\nu^\beta-\frac{q_\mu q_\nu}{q^2} 
\alpha_sG_{\rho\beta}G_\sigma^\beta q^\rho q^\sigma\right)\Pi^{G^2}_3(q^2)
+\alpha_sG_{\mu\rho}G_{\nu\sigma}q^\rho q^\sigma\Pi^{G^2}_4(q^2)
\\
&+\frac{q_\mu
q_\nu}{q^2}\alpha_sG_{\alpha\beta}G^{\alpha\beta}\Pi^{G^2}_5(q^2)
+\frac{q_\mu q_\nu}{q^2}\alpha_sG_{\rho\beta}G_\sigma^\beta q^\rho
q^\sigma
\Pi^{G^2}_6(q^2).
\end{split}
\end{equation} 
In order to extract $\langle\alpha_sG_{\mu\nu}G^{\mu\nu}\rangle$, we
need  a condition for the gluonic vacuum expectation value which reads 
\begin{equation}
G_{\mu\nu}G_{\rho\sigma}={\frac{1}{D(D-1)}
G_{\alpha\beta}G^{\alpha\beta}(g_{\mu\rho}g_{\nu\sigma}-
g_{\mu\sigma}g_{\nu\rho})}.
\end{equation}
Then, in  the $\rm\overline{MS}$-scheme and Feynman gauge, we obtain 
\begin{gather}\label{ga}
\Pi^{G^2a}_v(q^2)= {-
\frac{1}{36\pi}\ln\left(\frac{-q^2}{\mu^2}\right)\langle\alpha_sG^2\rangle
\left[\frac{16}{9}\frac{\alpha_s(\mu)}{\pi}\ln\left(\frac{-q^2}{\mu^2}\right)-
\frac{1139}{216}
\frac{\alpha_s(\mu)}{\pi}\right ]+{\rm infinite~terms}}\\
\Pi^{G^2a}_s(q^2)= {
\frac{1}{24\pi}\ln\left(\frac{-q^2}{\mu^2}\right)\langle\alpha_sG^2\rangle
\left[\frac{16}{9}\frac{\alpha_s(\mu)}{\pi}\ln\left(\frac{-q^2}{\mu^2}\right)-
\frac{1459}{216}
\frac{\alpha_s(\mu)}{\pi}\right ]+{\rm infinite~terms}}\quad ,
\end{gather}
where $\langle\alpha G^2\rangle\equiv\langle\alpha G^a_{\mu\nu}G^{a\,\mu\nu}\rangle$.
Another part of the  next-to-leading order calculation of
$\langle\alpha_sG_{\mu\nu}G^{\mu\nu}\rangle$ results  
from the renormalization of the current (\ref{z}), i.e.
\begin{equation}\label{1b} 
i\int \mathrm{d}^4x\, {\mathrm{e}}^{iqx} 2(Z-1)
\langle 0|T\{j _{\mu}(x),j^{+}_{\nu}(0)\}|0 \rangle,
\end{equation}
It reads
\begin{gather}\label{gb}
\Pi^{G^2b}_v(q^2)= {-
\frac{1}{36\pi}\ln\left(\frac{-q^2}{\mu^2}\right)\langle\alpha_sG^2\rangle
\left[-\frac{8\alpha_s(\mu)}{9\pi}\ln\left(\frac{-q^2}{\mu^2}\right)+\frac{88}{27}
\frac{\alpha_s(\mu)}{\pi}\right ]+{\rm infinite~terms}}\\
\Pi^{G^2b}_s(q^2)= {
\frac{1}{24\pi}\ln\left(\frac{-q^2}{\mu^2}\right)\langle\alpha_sG^2\rangle
\left[-\frac{8\alpha_s(\mu)}{9\pi}\ln\left(\frac{-q^2}{\mu^2}\right)+\frac{104}{27}
\frac{\alpha_s(\mu)}{\pi}\right ]+{\rm infinite~terms}},
\end{gather}
where the sum of the infinite terms in  (\ref{ga}) and 
(\ref{gb}) is scale-independent.  
We did not check on the infrared convergence of the sum of infinite terms because
we used  dimensional
regularization. However, the sum of these two parts must be  
IR convergent so that the Wilson coefficient of
the condensates only depends on short-distance effects. 
Obviously, this result is invariant if we use the background gauge,  
because all radiation fields are integrated out.   
We have checked such invariance.

\begin{figure}[hbt]
\centering
\includegraphics[scale=0.12]{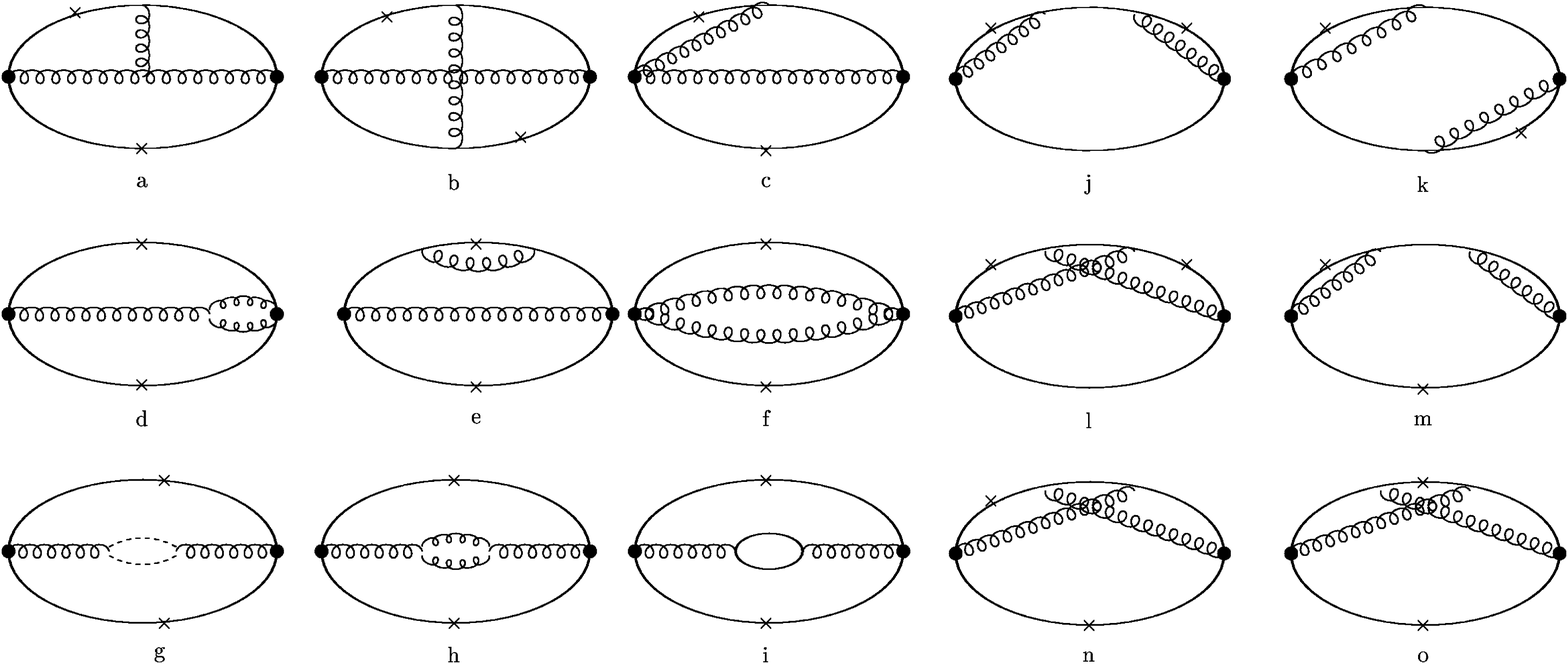}
\caption{
 Feynman diagrams for  the $\alpha_s$-corrections to the condensate$\langle
\bar qq\rangle^2$.
Dots stand for the current vertices.
}
\label{quark_fig}
\end{figure}

Similarly, 
by calculating the Feynman diagrams shown in Fig. \ref{quark_fig}, 
we  obtain the next-to-leading order corrections for the  $\langle\bar
qq\rangle^2$ contributions
\begin{gather}\label{q}
\Pi^{q}_v(q^2)= {\frac{4\pi}{9q^2}\alpha_s(\mu)\langle\bar
qq
\rangle^2\left[\left(\frac{11}{72}+\frac{1}{6}n_f\right)\frac{\alpha_s(\mu)}{\pi}
\ln\left(\frac{-q^2}{\mu^2}\right)+\left(\frac{91}{108}-\frac{5}{18}n_f\right)
\frac{\alpha_s(\mu)}{\pi}\right ]}\\
\Pi^{q}_s(q^2)={-\frac{4\pi}{3q^2}\alpha_s(\mu)\langle\bar
qq
\rangle^2
\left[\left(\frac{53}{72}+\frac{1}{6}n_f\right)\frac{\alpha_s(\mu)}{\pi}
\ln\left(\frac{-q^2}{\mu^2}\right)
-\left(\frac{14}{9}+\frac{1}{6}n_f\right)\frac{\alpha_s(\mu)}{\pi}\right ]}
\quad ,\end{gather}
where we have used the vacuum saturation approximation  
\begin{equation}
\langle\bar q_i^a q_j^b\bar q_{i^\prime}^{a^\prime}
q_{j^\prime}^{b^\prime}\rangle = {
(\frac{1}{12})^2\langle\bar qq\rangle^2
(\delta_{ij}\delta_{i^\prime j^\prime}\delta_{ab}\delta_{a^\prime 
b^\prime}
-\delta_{ij^\prime}\delta_{i^\prime j}\delta_{ab^\prime}\delta_{a^\prime
b})}\quad .
\end{equation}

\section{Mass of the $1^{-+}$  hybrid mesons}
QCD sum rules are  based on the resonance plus continuum duality ansatz
\begin{equation}
\frac{1}{\pi}Im\Pi_{v,s}(s)=\sum_R 
M_R^6f_R^2\delta\left(s-M_R^2\right)+{\rm QCD~continuum}, 
\end{equation}
where $M_R$ is the mass of the resonance $R$, $f_R$ denotes the
coupling of the resonance to the current and we use the narrow resonance
approximation $\delta(s-M_R^2)$ for the resonance $R$.
The spectral density $\rho_{v,s}(s)=\frac{1}{\pi}Im\Pi_{v,s}(s)$ can be 
related to the correlator $\Pi_{v,s}(q^2)$ at the scale $-q^2$ via the standard
dispersion relation 
\begin{equation}\label{dr}
\Pi_{v,s}(q^2)={(q^2)^n
\int^\infty_0ds\frac{\rho_{v,s}(s)}{s^n(s-q^2)}+\sum^{n-1}_{k=0}a_k(q^2)^k},
\end{equation}
where the $a_k$ are appropriate subtraction constants to render
Eq.\ (\ref{dr}) finite. The energy variable $-q^2$ has to be chosen in a region where 
one can incorporate the asymptotic freedom property of QCD via the
operator product expansion. The spectral function $\Pi_{v,s}(q^2)$ can be expressed as 
\begin{equation}
\Pi_{v,s}(q^2)=\sum_{dim~ n} C_n(-q^2,\mu){\it O}_n(\mu)
\end{equation}
where the scale $\mu$ separates the long-distance and short distance regime 
of QCD. The Wilson coefficients $C_n(-q^2,\mu)$ for the low dimension
operators were given in \cite{lead,jin} as summarized in Section 3 of this paper. 
Ref.\  \cite{narison}  also introduced a
dimension two operator resulting from the resummation  of the large order terms of the
OPE series. However, this method appears to be model-dependent and 
could result in a double counting of 
the contribution of the operators considered in the present approach.
We will therefore not include the dimension two 
term in our analysis.  

Concentrating on the analysis of the vector $1^{-+}$ channel, 
the lowest-lying resonance in the spectral density is enhanced by the standard approach of applying the
Borel transform operator to (\ref{dr}) weighted by powers of $q^2$ \cite{shif}. This results in the 
Laplace sum-rules 
\begin{equation}\label{bor}
R_k\left(\tau,s_0\right)=\int^{s_0}
s^k{\mathrm{e}}^{-s\tau} \rho_{v}(s)\mathrm{d}s~;~k=0,1,2,\ldots
\end{equation}
where the quantity $R_k$ represents the QCD prediction, and 
the threshold $s_0$ separates the contribution from 
higher excited states and the QCD continuum.
In the single narrow resonance scenario, the lowest-lying resonance mass can be obtained from  ratios
\begin{equation}
M_R^2=\frac{R_{k+1}\left(\tau,s_0\right)}{R_{k}\left(\tau,s_0\right)}\quad .
\end{equation}

The zero-weighted sum-rule $R_0$ for $n_f=3$ can be obtained from Eqs.\ (\ref{ga}) and (\ref{gb}), 
and Refs.\cite{lead,jin}, with some  results needed for calculation of the necessary Borel transforms 
extracted from \cite{glue}.  
\begin{equation}
\begin{split}
R_0\left(\tau,s_0\right)=&\frac{1}{240\pi^2}\frac{\alpha}{\pi}\left(1+\frac{1301}{240}\frac{\alpha}{\pi}\right)
\int\limits_0^{s_0}t^2 \mathrm{e}^{-t\tau}\mathrm{d}t 
-\frac{1}{120\pi^2}\frac{\alpha}{\pi}\left(\frac{17}{72}\frac{\alpha}{\pi}\right)
\int\limits_0^{s_0}t^2 \mathrm{e}^{-t\tau}\ln\left(\frac{t}{\mu^2}\right)\mathrm{d}t\\
&+\frac{1}{36\pi}\left(1-\frac{145}{72}\frac{\alpha}{\pi}\right)\langle\alpha G^2\rangle
\int\limits_0^{s_0} \mathrm{e}^{-t\tau}\mathrm{d}t 
+
\frac{1}{36\pi}\left(\frac{16}{9}\frac{\alpha}{\pi}\right)\langle\alpha G^2\rangle
\int\limits_0^{s_0} \mathrm{e}^{-t\tau}\ln\left(\frac{t}{\mu^2}\right)\mathrm{d}t \\
&-\langle {\cal O}_6\rangle-\frac{4\pi}{9}\left(1+\frac{1}{108}\frac{\alpha}{\pi} \right)
\alpha\langle\bar q q\rangle^2
-\frac{4\pi}{9}\left(\frac{47}{72}\frac{\alpha}{\pi} \right)\alpha\langle\bar q q\rangle^2
\left[-\gamma_E-\ln\left(\mu^2\tau\right)-E_1\left(s_0\tau\right)\right]
\end{split}
\label{R0_res}
\end{equation}
The quantity $\langle {\cal O}_6\rangle$ is defined by
\begin{equation}
\langle {\cal O}_6\rangle=\frac{1}{192\pi^2}\langle g^3G^3\rangle-\frac{83}{1728}\frac{\alpha}{\pi}
m_q\langle\bar q gGq\rangle\quad .
\end{equation}
Renormalization-group improvement of (\ref{R0_res}) is achieved by setting $\mu^2=1/\tau$ \cite{RG}, and  
higher-weight sum-rules can be obtained from $\tau$ derivatives of  (\ref{R0_res}) before implementing 
renormalization-group improvement. As stated earlier, this procedure 
implies that  the NLO $\alpha\langle\bar q q\rangle^2$ 
correction provides the leading contribution in $R_1$.

The  various QCD parameters that will be used in the phenomenological analysis of  (\ref{R0_res})  
are 
\begin{gather}\label{par}
\Lambda_{{\overline{MS}}} \approx 0.3\, {\rm GeV}~,~ \alpha_s\langle\bar qq\rangle^2 =
1.8\times 10^{-4}\,{\rm GeV^6}~,~m_q\langle\bar qq\rangle
=-\frac{1}{4}f_\pi^2m_\pi^2\\
\langle\alpha_sG^2\rangle=0.07\,{\rm GeV^4}~,~g^3\langle G^3\rangle =
1.1\,{\rm GeV^2}\langle\alpha_sG^2\rangle~,~ f_\pi=0.132\,{\rm GeV}
~,~m_q\langle\bar qgGq\rangle=1.5\,{\rm GeV^2}m_q\langle\bar
qq\rangle.
\label{more_par}
\end{gather}
The parameter $\langle\alpha_sG^2\rangle$ represents the central value in the recent determination  \cite{nar1},
$g^3\langle G^3\rangle$ is obtained from the dilute instanton gas model \cite{dilute}, 
$\langle\bar qgGq\rangle$ is extracted from \cite{reind}, and the dimension-six condensate
parameter $\alpha_s\langle\bar qq\rangle^2$ is referenced to the vacuum saturation value which is 
known to underestimate the actual value by up to a factor of 2 in the ($I=1$) vector and axial vector channels \cite{saturation}.

\begin{figure}[hbt]
\centering
\includegraphics[scale=0.3]{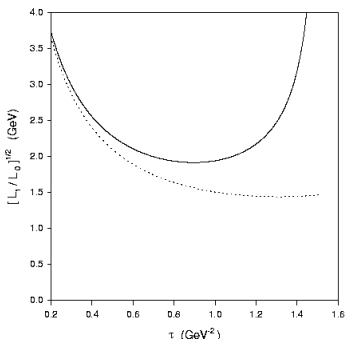}
\caption{
The ratio $\sqrt{R_1\left(\tau,\infty\right)/R_0\left(\tau,\infty\right)}$ as a function of $\tau$
for the NLO (solid curve) and LO (dashed curve) sum-rules.  As discussed in the text, acceptable values of $M_R$ must lie below these curves.
}
\label{bounds_fig}
\end{figure}

Before considering a detailed analysis of the sum-rules, we consider the $s_0\to\infty$ limit 
of the sum-rules which provides the following  bound on the lightest resonance mass
\begin{equation}
M_R^2\le\frac{R_{1}\left(\tau,\infty\right)}{R_{0}\left(\tau,\infty\right)}\quad ,
\end{equation}
which has the advantage of being a robust bound independent of the QCD continuum 
model.  The explicit expressions for the
sum-rules in the  $s_0\to\infty$ limit are
\begin{gather}
\begin{split}
R_{0}\left(\tau,\infty\right)=&
\frac{1}{240\pi^2}\frac{\alpha}{\pi}\left(1+\frac{1301}{240}\frac{\alpha}{\pi}\right)\frac{2}{\tau^3}
-\frac{1}{120\pi^2}\left(\frac{\alpha}{\pi}\right)^2\frac{17}{72}\left(3-2\gamma_E\right)\frac{1}{\tau^3}
+\frac{1}{36\pi}\left(1-\frac{145}{72}\frac{\alpha}{\pi}\right)\langle\alpha G^2\rangle\frac{1}{\tau}
\\
&-\frac{1}{36\pi}\frac{16}{9}\frac{\alpha}{\pi}\langle\alpha G^2\rangle\frac{\gamma_E}{\tau}
-\langle {\cal O}_6\rangle-\frac{4\pi}{9}\left(1+\frac{1}{108}\frac{\alpha}{\pi} \right)
\alpha\langle\bar q q\rangle^2
+\frac{4\pi}{9}\left(\frac{47}{72}\frac{\alpha}{\pi} \right)\alpha\langle\bar q q\rangle^2\gamma_E
\end{split}
\\
\begin{split}
R_{1}\left(\tau,\infty\right)=&
\frac{1}{240\pi^2}\frac{\alpha}{\pi}\left(1+\frac{1301}{240}\frac{\alpha}{\pi}\right)\frac{6}{\tau^4}
-\frac{1}{120\pi^2}\left(\frac{\alpha}{\pi}\right)^2\frac{17}{72}\left(11-6\gamma_E\right)\frac{1}{\tau^4}
\\
&+\frac{1}{36\pi}\left(1-\frac{145}{72}\frac{\alpha}{\pi}\right)\langle\alpha G^2\rangle\frac{1}{\tau^2}
-\frac{1}{36\pi}\frac{16}{9}\frac{\alpha}{\pi}\langle\alpha G^2\rangle\frac{1-\gamma_E}{\tau^2}
-\frac{4\pi}{9}\left(\frac{47}{72}\frac{\alpha}{\pi} \right)\alpha\langle\bar q q\rangle^2\frac{1}{\tau}\quad .
\end{split}
\end{gather}
The effect of the NLO $\langle \alpha G^2\rangle$ and $\alpha\langle \bar q q\rangle^2$ 
corrections is illustrated in Figure \ref{bounds_fig}, where it is observed that the
mass bound is increased significantly when these higher-order corrections are included. 
For brevity, we  respectively refer to sum-rules 
containing the NLO and LO $\langle \alpha G^2\rangle$ and $\alpha\langle \bar q q\rangle^2$ 
corrections as the NLO  and LO sum-rules.  
In particular, 
we see from Fig.\ \ref{bounds_fig} that the $\hat\rho(1600)$, excluded  for the LO case,  can be accommodated when the NLO corrections 
are included.  Furthermore, the minimum of the NLO bounds occurs at a reasonable energy ($\tau$) scale 
in contrast to the rather large  energy scale occurring when only LO corrections are included.

\begin{figure}[hbt]
\centering
\includegraphics[scale=0.3]{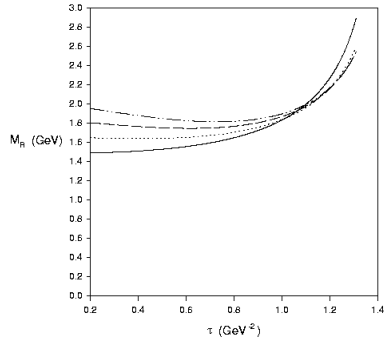}
\caption{
The NLO sum-rule ratio $\sqrt{R_1\left(\tau,s_0\right)/R_0\left(\tau,s_0\right)}$ as a function of $\tau$
for the sequence of $s_0$ values $s_0=\left\{3.0,~4.0,~5.0,~6.0\right\}\,{\rm GeV^2}$. The lowest (solid) curve corresponds to 
$s_0=3.0\,{\rm GeV^2}$, and the upper (dashed-dotted) curve corresponds to   $s_0=6.0\,{\rm GeV^2}$. 
}
\label{NLO_mass_fig}
\end{figure}

\begin{figure}[hbt]
\centering
\includegraphics[scale=0.3]{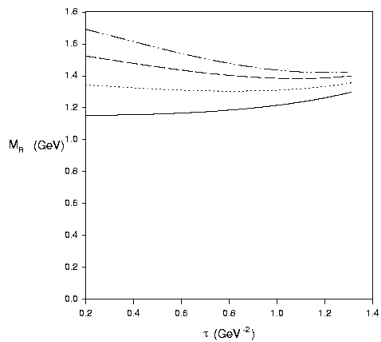}
\caption{
The LO sum-rule ratio $\sqrt{R_1\left(\tau,s_0\right)/R_0\left(\tau,s_0\right)}$ as a function of $\tau$
for the sequence of $s_0$ values $s_0=\{2.0~,3.0,~4.0,~5.0\}\,{\rm GeV^2}$. The lowest (solid) curve corresponds to 
$s_0=2.0\,{\rm GeV^2}$, and the upper (dashed-dotted) curve corresponds to   $s_0=6.0\,{\rm GeV^2}$. 
}
\label{LO_mass_fig}
\end{figure}

The mass estimate $M_R$ is obtained by optimizing the choice of $s_0$ such that the most stable $R_1/R_0$ ratio is obtained.  
Figures \ref{NLO_mass_fig} and \ref{LO_mass_fig} illustrate this ratio for selected values of $s_0$,   
resulting in  $M_R\approx 1.6\,{\rm GeV}$ and $s_0\approx 4.0\,{\rm GeV^2}$    for the NLO case, while the LO analysis
results in $M_R\approx 1.3\,{\rm GeV}$  and $s_0\approx 3.0\,{\rm GeV^2}$. These optimized values of $M_R$ 
are consistent with the bounds established in Fig.\ \ref{bounds_fig}, and explicitly demonstrate that the NLO 
condensate effects raise the estimated value of the hybrid mass.

\begin{figure}[hbt]
\centering
\includegraphics[scale=0.3]{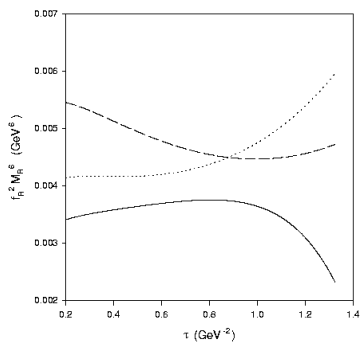}
\caption{
The quantity $M_R^{-2k}\mathrm{e}^{-M_R^2\tau}R_k\left(\tau,s_0\right)$, which in the single-resonance model
corresponds to $f_R^2M_R^6$, is displayed as a function of $\tau$ for the $k=0,1,2$ NLO sum-rules.  The optimized values 
$s_0=4.0\,{\rm GeV^2}$ and $M_R=1.64\,{\rm GeV}$ are used as inputs, the lowest (solid) curve corresponds to $k=0$,
the intermediate (dotted) curve corresponds to $k=1$, and the upper (dashed) curve corresponds to $k=2$.
}
\label{NLO_f_fig}
\end{figure}

\begin{figure}[hbt]
\centering
\includegraphics[scale=0.3]{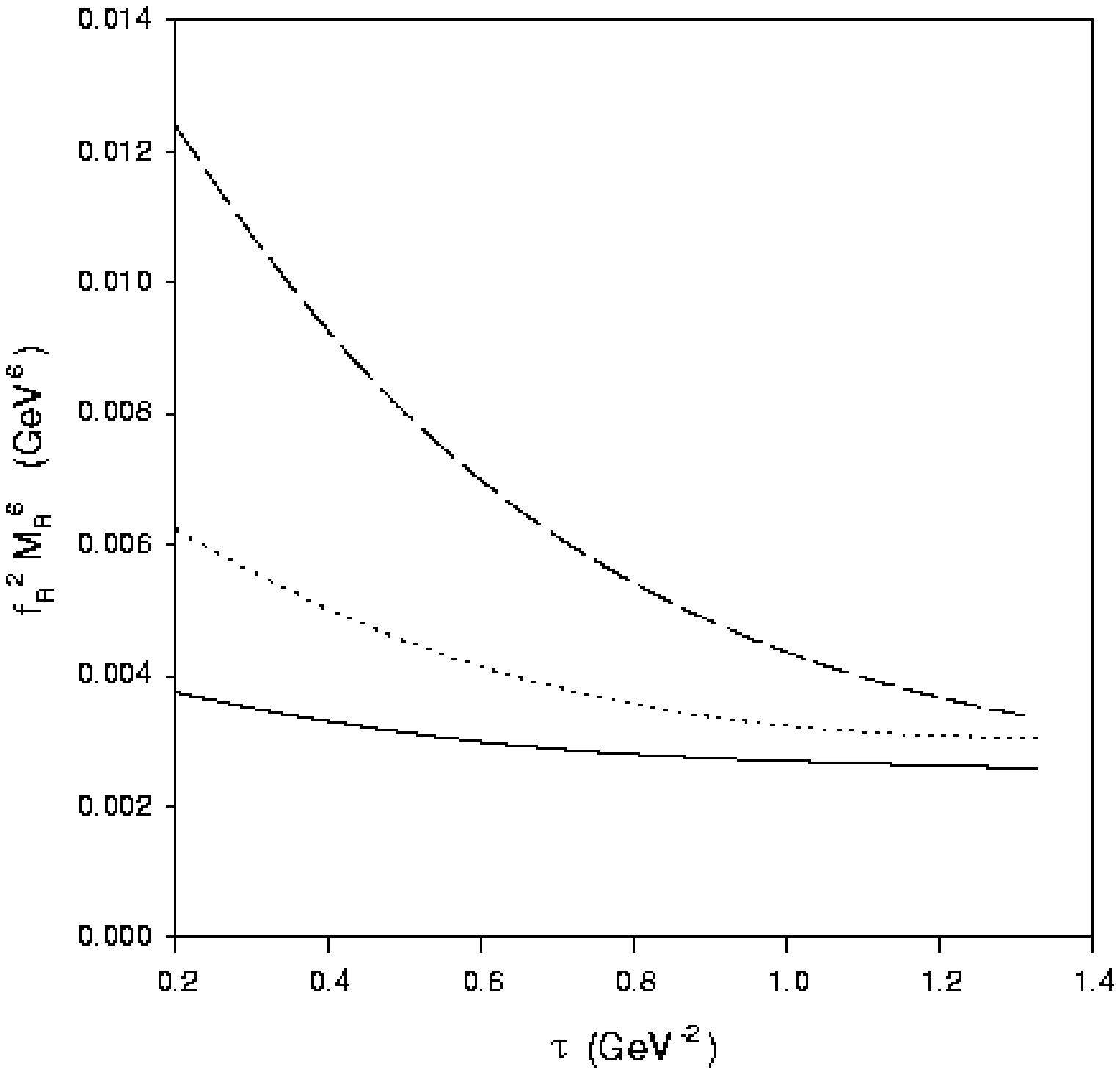}
\caption{
The quantity $M_R^{-2k}\mathrm{e}^{-M_R^2\tau}R_k\left(\tau,s_0\right)$, which in the single-resonance model
corresponds to $f_R^2M_R^6$, is displayed as a function of $\tau$ for the  $k=0,1,2$ LO sum-rules.  
The optimized values 
$s_0=3.0\,{\rm GeV^2}$ and $M_R=1.31\,{\rm GeV}$ are used as inputs, the lowest (solid) curve corresponds to $k=0$,
the intermediate (dotted) curve corresponds to $k=1$, and the upper (dashed) curve corresponds to $k=2$.
}
\label{LO_f_fig}
\end{figure}

The stability and self-consistency of the sum-rule analyses can be examined by using the optimized $s_0$ and $M_R$ as 
input into the following expressions resulting from the single resonance model.
\begin{equation}
\frac{\mathrm{e}^{-M_R^2\tau}}{M_R^{2k}}R_k\left(\tau,s_0\right)=f_R^2M_R^6
\label{sr_family}
\end{equation}
Figure \ref{NLO_f_fig} displays the left-hand side of (\ref{sr_family}) for the NLO $k=0,1,2$ sum-rules.  
We see from Fig.\ \ref{NLO_f_fig} that
a stable ratio containing a $\tau$ critical point occurs for each value of $k$, and that the variation of the corresponding value of $f_R^2M_R^6$   
with $k$ is minimal.  By contrast, the corresponding curves for the LO sum-rules shown in 
Figure \ref{LO_f_fig} do not exhibit a critical point in the  same $\tau$ range associated with the 
Fig.\ \ref{LO_mass_fig} mass estimate, and show strong dependence on $k$.  Thus the NLO
 $\langle \alpha G^2\rangle$ and $\alpha\langle \bar q q\rangle^2$ 
corrections lead to improved stability and self-consistency in the sum-rule analysis.

The dominant uncertainties in the sum-rule analysis associated with the QCD parameters 
(\ref{par}) and (\ref{more_par}) are related to $\Lambda_{\overline{{\rm MS}}}$ and $\alpha\langle\bar qq\rangle^2$.   If the
pattern established in other sum-rule channels \cite{saturation} is upheld for the hybrid OPE, then 
$\alpha\langle\bar qq\rangle^2$  given in (\ref{par}) underestimates the true value.  Similarly, 
 $\Lambda_{\overline{{\rm MS}}}=300\,{\rm MeV}$  is a lower bound following from $\alpha_s\left(M_\tau\right)$ \cite{pdg}.
Increasing either of these parameters increases the hybrid mass estimates, and thus it appears difficult to accommodate the $\hat\rho(1400)$.

Finally, we have verified that the results of our analysis are essentially 
independent of the choice of renormalization scale $\mu^2=1/\tau$ motivated by
renormalization-group improvement in the $s_0\to\infty$ limit \cite{RG}.  
Choosing renormalization scales in the energy region near the hybrid mass 
has a minimal effect on the sum-rule analysis.

\section{Summary}
The (NLO)  $\alpha_s$ corrections to 
the   $\langle\alpha_s G^2\rangle$ and  $\alpha\langle\bar qq\rangle^2$ 
contributions in the  two point  correlator of the current 
$g\bar q\gamma_{\nu}iG_{\mu\nu}^aT^aq(x)$ have been calculated, and the effect of these contributions on the QCD sum-rule estimates of the
 $1^{-+}$  hybrid  mass 
have been examined.  The NLO $\alpha\langle\bar qq\rangle^2$ corrections are particularly interesting since they provide the {\em leading} 
contributions to the $R_1$ sum-rule.  The NLO  $\langle\alpha_s G^2\rangle$ and  $\alpha\langle\bar qq\rangle^2$  contributions improve the stability and 
self-consistency of the sum-rule analysis, resulting in a $1^{-+}$  hybrid  mass of approximately $1.6\,{\rm GeV}$.  This result reflects a lower 
bound devolving from the QCD input parameters, so it appears difficult to accommodate the $\hat\rho(1400)$ as a hybrid state.

\begin{acknowledgments}
H.Y.~Jin  would like to thank  A. Pivovarov for very useful
discussions. He also thanks the University of Saskatchewan for its warm 
hospitality.  The work of H.Y. Jin is  supported  by NSCF. This work was   begun
while H.Y.~Jin was a fellow of the Alexander-von-Humboldt foundation at the
University of Mainz. He would like to thank the AvH foundation for support
and the University of Mainz for its hospitality.   TGS gratefully acknowledges research support from the 
Natural Sciences \& Engineering Research Council of Canada.
\end{acknowledgments}

\end{document}